# Scattering of a few-cycle laser pulse by a plasma layer : the role of the carrier-envelope phase difference at relativistic intensities

Sándor Varró

*Research Institute for Solid State Physics and Optics*
*Letters : H-1525 Budapest, POBox 49, Hungary*
*E-mail : varro@sunserv.kfki.hu*

**Abstract.** The reflection and transmission of a few-cycle femtosecond Ti:Sa laser pulse impinging on a thin plasma layer have been analysed on the basis of classical electrodynamics. The plasma electrons were represented by a surface current density along the layer. This target plasma can be imagined as generated from a thin foil by a pre-pulse which is then followed by the main high-intensity laser pulse. An approximate analytic solution has been given for the system of the coupled Maxwell-Lorentz equations describing the dynamics of the surface current and the composite radiation field. With the help of these solutions the Fourier components of the reflected radiation have been calculated. The nonlinearities stemming from the relativistic kinematics of the free electrons lead to the appearance of higher-order harmonics in the scattered spectra. In our analysis particular attention has been paid to the role of the carrier-envelope phase difference of the incoming few-cycle laser pulse. For instance, we show that the fourth harmonic peak strongly depends on the carrier-envelope phase difference with a modulation being almost 25 percent. In general, the harmonic peaks are down-shifted due to the presence of the intensity-dependent factors of order of 15-65 percent in case of an incoming field of intensity $2 \times 10^{19}$ W/cm$^2$. This phenomenon is analogous to the famous intensity-dependent frequency shift appearing in the high-intensity Thomson scattering on a single electron. The analysis has shown that at grazing incidence the first even harmonics dominate rather than the odd ones, as for medium angles of incidence. It has also been shown that the spectrum of the high-harmonics has a long tail where the heights of the peaks vary practically within one order of magnitude forming a quasi-continuum. By Fourier synthesising the components from this plateau region of the higher-harmonic spectrum, attosecond pulses have been obtained.

## 1. Introduction

The study of the interaction of intense few-cycle laser pulses with matter has brought a new, important branch of investigations in nonlinear optics, as Brabec and Krausz [1] emphasized in their review paper. The effect of the absolute phase (the carrier-envelope phase difference, in short: CE phase) on the nonlinear response of atoms and of solids interacting with a very short, few-cycle strong laser pulse has recently drawn considerable attention and initiated a wide-spreading theoretical and experimental research. For instance Paulus et al. [2] have detected an anticorrelation in the shot-to-shot analysis of the photoelectron yield of ionization measured by two opposing detectors. This effect comes from the random variation of the CE phase (hence the direction and the magnitude of the electric field of the laser) from one pulse to the other. Such extreme short pulses can be used to monitor for instance the details of photoelectron dynamics [3] or atomic inner-shell relaxation processes, like the Auger effect [4]. Concerning theory, the CE pase-dependence of the spatial asymmetry in photoionization has been extensively investigated by Chelkowski et al. [5], and by Milosevic et al. [6-7]. In the meantime the problem of the stabilization of the CE phase in the few-cycle laser pulse trains has been achieved [8-9]. On the basis of a simulation, using the time-dependent density functional approach, Lemell et al. [10] predicted the CE phase-dependence of the photoelectron yield in case of the surface photoelectric effect of metals in the tunnel regime. Apolonskiy et al. [11] and Dombi et. al. [12] have reported the measurement of this effect, but the absolut phase-dependence had a considerably smaller modulation (CE phase-sensitivity) in their experiment, than predicted by the simulation. Fortier et al. has recently demonstrated the CE phase effect in quantum interference of injected photocurrents in semiconductors [13]. In the multiphoton regime Nakajima and Watanabe [14] has found theoretically CE phase effects in the bound state population of a Cs atom excited by nearly single-cycle pulses.

At this point we would like to note that there is a wide-spread opinion among researchers investigating the CE phase effect that this effect shows up exclusively in *nonlinear* processes. In fact, as Fearn and Lamb [15] have shown already in 1991, the sine or the cosine character of the laser pulse make a difference in the *linear* photoionization dynamics if one takes into account the counter-rotating term in the interaction. As they wrote in Section IV. of their paper: "This suggests that…the time delay [of the electron signal] could be used to measure the phase of the field." A simple illustration of the linear CE phase effect has recently been considered by Ristow [16] in the case of a harmonic oscillator.

We have seen above, that in the theoretical works exclusively nonlinear quantum processes (photoionization, surface photoelectric effect) have been considered in the non-relativistic regime in the context of CE phase effects. In the present paper we briefly describe our theoretical analysis on the reflection and transmission of a few-cycle laser pulse on a thin plasma layer represented by a surface current. Our analysis here, as in our earlier non-relativistic study [17] on CE phase effects in high-harmonic generation on a thin metal layer, is based completely on classical electrodynamics and mechanics, in the frame of which we solve the system of coupled Maxwell-Lorentz equations of the incoming and scattered radiation and the classical surface current. The idea to study such a system appeared to us by reading a paper by Sommerfeld [18] published in Annalen der Physik in 1915, in which he analysed the temporal distortion of x-ray pulses of arbitrary shape and duration impinging perpendicularly on a surface current being in vacuum. In Ref. [17] we have generalized this model in the following sense. On one hand, we allowed oblique incidence of the incoming radiation field, and on the other hand, we assume that the surface current (which represents a thin metal layer) is embedded between two semi-infinite dielectrics with two different indeces of refraction. In the present paper we shall give the relativistic generalization of the equation of motion of the surface current, which was not investigated by Sommerfeld. This latter approach is needed to consider the interaction of a plasma layer with a laser pulse of relativistic intensities. In our analysis we shall use the simplifying assumption that the plasma layer surrouded by vacuum, hence the angle of incidence coincides the angle of reflection and of transmission. Moreover, the thickness of the layer will be assumed to be smaller than the wavelength of the incoming radiation. This assumptuon restricts the applicability of our results to moderate relativistic intensities (for which the intensity parameter is not larger than unity), because for higher intensities the displacements of the plasma electrons along the cross-section of the layer can be larger than the wavelength of the incoming radiation.

The high-harmonic production on metallic free electrons has been considered analytically by Lichters et al. [19] on the basis of their famous oscillating mirror model. Our model considerably differs from this approach, because we have exactly taken into account the radiation reaction on the electron plasma. We note that recently there has been much labour put into the classical simulations of various processes (generation on coherent x-rays, laser acceleration of electrons) in laser-plasma interaction [20-22]. Moreover, this year the first experimental results appeared on the generation of quasi-monoenergetic electron bunches by strong laser fields [23]. On the other hand, to our knowledge, there have been no classical relativistic considerations so far published, where the CE phase effect in the relativistic regime would have been analysed. The recent paper may be considered as a contribution to the study of this particular aspect, too.

In Section 2 we present the basic equations describing our model. In Section 3 an approximate analytic solution will be given for the system of the coupled Maxwell-Lorentz equations describing the dynamics of the surface current (representing the plasma electrons) and the composite radiation field. With the help of these solutions the Fourier components of the reflected and transmitted radiation will be calculated. The nonlinearities stemming from the relativistic kinematics lead to the appearance of higher-order harmonics in the scattered spectra. In general, the harmonic peaks are down-shifted due to the presence of the intensity-dependent factors of order of 15-65 per cent in case of an incoming field of intensity $2 \times 10^{19}$ W/cm$^2$. This phenomenon is analogous to the famous intensity-dependent frequency shift appearing in the high-intensity Thomson scattering on a single electron. In our analysis particular attention has been paid to the role of the carrier-envelope phase difference of the incoming few-cycle laser pulse. For instance the 4$^{th}$ harmonic peak strongly depends on the carrier-envelope phase difference with a modulation being almost 25 per cent. It is also shown that the scattered spectrum has a long tail, where the heights of the peaks vary practically within one order of magnitude in the frequency range considered. We will show that by Fourier synthesising the components from this plateau region of the higher-harmonic spectrum, attosecond pulses can be obtained.

**2. The basic equations of the model**

The model to be used here has already been studied in our earlier work [17]. For completeness of the present paper, let us briefly summarize first the basic notations and equations, which can also be found in Ref. [17] for the general case, when the electron layer is embedded between two dielectrics. Later we will specialize our study to case of an electron (plasma) layer being in vacuum. We take the coordinate system such that the first dielectric with index of refraction $n_1$ fills the region $z > l_2/2$, this is called region 1. In region 2 we place the thin metal layer of thickness $l_2$ perpendicular to the z-axis and defined by the relation $-l_2/2 < z < +l_2/2$. Region 3, $z < -l_2/2$, is assumed to be filled by the second dielectric having the index of refraction $n_3$. The thickness $l_2$ is assumed to smaller then the wavelength of the incoming radiation. The plasma layer is represented by a sheet of electrons bound to region 2 and moving freely in the x-y plane. In case of perpendicular incidence the light would come from the positive z-direction, and it would be transmitted in the negative z-direction into region 3. The plane of incidence is defined as the y-z plane and the initial $\bar{k}$-vector is assumed to make an angle $\theta_1$ with the z-axis. In case of an s-polarized incoming TE wave the components of the electric field and the magnetic induction read $(E_x,0,0)$ and $(0, B_y, B_z)$, respectively. They satisfy the Maxwell equations

$$\partial_y B_z - \partial_z B_y = \partial_0 \varepsilon E_x, \quad \partial_z E_x = -\partial_0 B_y, \quad -\partial_y E_x = -\partial_0 B_z \quad , \tag{1}$$

where $\varepsilon = n^2$ is the dielectric constant and $n$ is the index of refraction. If we make the replacements $\varepsilon E_x \to -B_x$, $B_z \to E_z$ and $B_y \to E_y$ then we have the field components of a p-polarized TM wave $(0, E_y, E_z)$ and $(B_x, 0, 0)$, and we receive the following equations

$$\partial_z B_x = \partial_0 \varepsilon E_y, \quad -\partial_y B_x = \partial_0 \varepsilon E_z, \quad \partial_y E_z - \partial_z E_y = -\partial_0 B_x \quad . \tag{2}$$

In the followings we will consider only the latter case, namely the scattering of a p-polarized TM radiation field. From Eq. (2) we deduce that $B_x$ satisfies the wave equation, and in region 1 we take it as a superposition of the incoming plane wave pulse $F$ and an unknown reflected plane wave $f_1$

$$B_{x1} = F - f_1 = F[t - n_1(y\sin\theta_1 - z\cos\theta_1)/c] - f_1[t - n_1(y\sin\theta_1 + z\cos\theta_1)/c]. \quad (3)$$

From Eq.(2) we can express the components $E_y$ and $E_z$ of the electric field strength by taking into account Eq. (3)

$$E_{y1} = (\cos\theta_1/n_1)(F + f_1), \quad E_{z1} = (\sin\theta_1/n_1)(F - f_1). \quad (4)$$

In region 3 the general form of the magnetic induction $B_{x3}$ is the by now unknown refracted wave $g_3$

$$B_{x3} = g_3 = g_3[t - n_3(y\sin\theta_3 - z\cos\theta_3)/c]. \quad (5)$$

The corresponding components of the electric field strength are expressed from the above equation with the help of the first two equation of Eq. (2)

$$E_{y3} = (\cos\theta_3/n_3)g_3, \quad E_{z3} = (\sin\theta_3/n_3)g_3. \quad (6)$$

In region 2 the relevant Maxwell equations with the current density $\vec{j}$ read

$$\partial_z B_x = (4\pi/c)j_{y2} + \partial_0 \varepsilon E_y, \quad \partial_y E_z - \partial_z E_y = -\partial_0 B_x. \quad (7)$$

By integrating the two equations in Eq. (7) with respect to z from $-l_2/2$ to $+l_2/2$ and taking the limit $l_2 \to 0$, we obtain the boundary conditions for the field components

$$[B_{x1} - B_{x3}]_{z=0} = (4\pi/c)K_{y2}, \quad [E_{y1} - E_{y3}]_{z=0} = 0, \quad (8)$$

where $K_{y2}$ is the y-component of the surface current in region 2. This surface current can be expressed in terms of the local velocity of the electrons in the metal layer

$$K_{y2} = e(d\delta_y/dt)l_2 n_e, \quad (4\pi/2c)K_{y2} = (m/e)\Gamma(d\delta_y/dt), \quad (9)$$

$$\Gamma \equiv 2\pi(e^2/mc)l_2 n_e, \quad \Gamma = (\omega_p/\omega_0)^2(\pi l_2/\lambda_0)\omega_0, \quad \kappa \equiv \Gamma/\omega_0 = \pi(\omega_p/\omega_0)^2(l_2/\lambda_0), \quad (10)$$

where for later convenience we have introduced $\omega_0$, $\lambda_0 = 2\pi c/\omega_0$, the carrier frequency and the central wavelength of the incoming light pulse, and $n_e, \omega_p = \sqrt{4\pi n_e e^2/m}$ denote the density of electrons and the corresponding plasma frequency in the metal layer, respectively. In Eq. (9) $\delta_y$ denotes the local displacement of the electrons in the metal layer for which we later write down the Lorentz equation (Newton equation in the non-relativistic regime) in the presence of the complete electric field. We remark that in reality the thickness $l_2$ is, of course, not infinitesimally small, rather, it has a finite value which is anyway assumed to be smaller then the average wavelength of the incoming pulse.

From Eq. (8) with the help of Eq. (9) we can express $f_1$ and $g_3$ in terms of $\delta_y'(t')$

$$f_1(t') = (1/(c_1 + c_3))[(c_3 - c_1)F(t') - 2c_3(m/e)\Gamma\delta_y'(t')], \quad (11)$$

$$g_3(t') = (2c_1/(c_1 + c_3))[F(t') - (m/e)\Gamma\delta_y'(t')], \quad (12)$$

where the prime on $\delta_y$ denotes the derivative with respect to the retarded time $t' = t - yn_1\sin\theta_1/c$ which is equal to $t - yn_3\sin\theta_3/c$, securing Snell's law of refraction $n_1\sin\theta_1 = n_3\sin\theta_3$ to hold. Moreover, in Eqs. (11) and (12) we have introduced the notations $c_1 = \cos\theta_1/n_1$, $c_3 = \cos\theta_3/n_3$. We would like to emphasize that Eqs. (11) and (12) are valid in complete generality, that is, they hold for both non-relativistic and relativistic kinematics of the local electron displacement $\delta_y(t')$. For an interaction with a TM wave this displacement is uniform (along lines of constant x-values) in the direction perpendicular to the plane of incidence (the *y-z* plane), so it does not depend on the *x*-coordinate. As the incoming wave impinges on the surface at region 2 its (plane) wave fronts sweep this surface creating a superluminar polarization wave, described by the local displacement $\delta_y(t')$ of the electrons. In the non-relativistic regime, because of the continuity of $E_y$, Eq. (8), in the Newton equation for the displacement of the electrons in the surface current we can use for instance the force term $E_{y1} = c_1(F + f_1)$ according to Eq. (4), and neglect the magnetic induction. By taking Eq. (11) also into account we have

$$\delta_y''(t') = b[(e/m)F(t') - \Gamma \delta_y'(t')], \tag{13}$$

where $b \equiv 2c_1c_3/(c_1+c_3)$, and $c_1 \equiv \cos\theta_1/n_1$, $c_3 \equiv \cos\theta_3/n_3$.

Henceforth, in the equation of motion for the surface current $K_{y2}$ we use $B_{x3}$, $E_{y3}$ and $E_{z3}$, moreover we specialize our system to be a plasma layer in vacuum (hence $n_1 = n_3 = 1 \rightarrow \theta_1 = \theta_3 = \theta$, consequently $c_1 = c_2 = \cos\theta$ and $b=1$). If we take into account retardation, i.e. relativistic effects, the argument of the field strength will be given in the layer as

$$\eta \equiv \left[t - \frac{(y+\delta_y)\sin\theta - (z+\delta_z)\cos\theta}{c}\right]_{z=0}. \tag{14}$$

The local displacements $\delta_y$ and $\delta_z$ in the layer will depend on the retarded time $t' = t - y\sin\theta/c$. In fact, these displacements represent a polarization wave in the layer propagating in the positive *y*-direction with the velocity $c/\sin\theta$. This is due to the assumed oblique incidence of the incoming field, i.e. the wave front of the incoming field sweeps the layer with such a velocity. The fields to be used in the equation of motion of the electron displacements can be expressed as $\vec{E} = \vec{\varepsilon}g_3$ and $\vec{B} = g_3(1,0,0) = \vec{n} \times \vec{E}$, where $\vec{\varepsilon} = (0,\cos\theta,\sin\theta)$ is the polarization vector and $\vec{n} = (0,\sin\theta,-\cos\theta)$ is the unit propagation vector. The true retarded time parameter is given from eq(14) as $\eta = t' - \vec{n} \cdot \vec{\delta}/c$ where $\vec{\delta} = (0,\delta_y,\delta_z)$. Introducing the velocity $\vec{v} = d\vec{\delta}/dt'$ and the associated relativistic factor $\gamma = 1/\sqrt{1-v^2/c^2}$ we can define the proper time element $d\tau = dt'/\gamma$. Moreover we define the four-position $\delta^\mu = \{ct',\vec{\delta}\}$ and the four-velocity $u^\mu = d\delta^\mu/d\tau = \{u_0,\vec{u}\}$. In this way the following set of relativistic equations derives for the four-velocity associated to the displacements in the plasma layer

$$m(d\vec{u}/d\tau) = (e/c)(u_0\vec{E} + \vec{u}\times\vec{B}), \text{ and } m(du_0/d\tau) = (e/c)\vec{u}\cdot\vec{E}. \tag{15}$$

We stress that in Eq. (15) the field strengths depend on the true retarded time parameter $\eta = t' - \vec{n}\cdot\vec{\delta}/c$ at the position of the electrons, where $t' = t - y\sin\theta/c$ is the uniform retarded time over the plasma layer. We note that, because of the assumed geometry of the scattering, the electrons move collectively in phase along the lines parallel to the *x*-axis. The second of the two equations in Eq. (15) expresses the relativistic work theorem. From the two equations in Eq. (15) the important relations can be derived,

$$u_0 - \vec{n}\cdot\vec{u} = ca = const., \tag{16}$$

and, as a consequence, on the other hand

$$cd\eta/d\tau = \frac{d}{d\tau}[ct' - \vec{n}\cdot\vec{\delta}(t')] = u_0 - \vec{n}\cdot\vec{u} = ca, \quad d\eta/d\tau = a = \gamma(1-\vec{n}\cdot\vec{v}/c), \tag{17}$$

where *a* is a constant depending on the initial local velocity. This means, that the derivatives with respect to the proper time are proportional to the derivatives with respect to the argument of the field strengths, $d/d\tau = ad/d\eta$, where the constant *a* depends only on the initial conditions. As is seen from the last relation in Eq. (17), for a particle initially at rest this constant is $a = 1$. According to the relation $d/d\tau = a\, d/d\eta$, the equation of motion in Eq. (15) can be brought to the form

$$\frac{d^2\vec{\delta}}{d\eta^2} = \frac{e}{ma}\left[\vec{E} + \vec{n}\left(\frac{1}{c}\frac{d\vec{\delta}}{d\eta}\cdot\vec{E}\right)\right], \qquad (18)$$

where, as we saw before $\vec{E} = \vec{\varepsilon}g_3$, and, according to Eq. (12)

$$g_3 = F - (m/e)\Gamma(a/\gamma)(d\delta_y/d\eta). \qquad (19)$$

Now let us make the following decomposition of the displacement $\vec{\delta}$,

$$\delta_\perp = \vec{\varepsilon}\cdot\vec{\delta} = \delta_y\cos\theta + \delta_z\sin\theta, \quad \delta_\parallel = \vec{n}\cdot\vec{\delta} = \delta_y\sin\theta - \delta_z\cos\theta, \qquad (20)$$

that is

$$\delta_y = \delta_\perp\cos\theta + \delta_\parallel\sin\theta, \quad \delta_z = \delta_\perp\sin\theta - \delta_\parallel\cos\theta. \qquad (21)$$

With the help of this decomposition and by integration with respect to $\eta$, it can be derived from Eq. (18) that

$$d\delta_\parallel/d\eta = \frac{1}{2c}[(d\delta_\perp/d\eta)^2 - (d\delta_\perp/d\eta)_0^2]. \qquad (22)$$

Henceforth we will take the initial value $(d\delta_\perp/d\eta)_0 = 0$ which corresponds to $a = 1$. In this way we obtain

$$d\delta_y/d\eta = (d\delta_\perp/d\eta)\cos\theta + (1/2c)(d\delta_\perp/d\eta)^2\sin\theta. \qquad (23)$$

Thus, if we solve the equation of motion for $d\delta_\perp/d\eta$, we can express $d\delta_y/d\eta$ through which, according to Eq. (19), the transmitted field can be calculated. Similarily, because of Eq.(11), the reflected field can also be determined.

Combining Eqs. (19) and (23) – after some lengthy but straightforward algebra – we receive the following closed equation for $\delta_\perp$

$$\frac{d^2\delta_\perp}{d\eta^2} = \frac{e}{m}F(\eta) - \Gamma\frac{(d\delta_\perp/d\eta)\cos\theta + [(d\delta_\perp/d\eta)^2/2c]\sin\theta}{\sqrt{1 + (d\delta_\perp/d\eta)^2/c^2 + (d\delta_\perp/d\eta)^4/4c^4}}. \qquad (24)$$

In case of perpendicular incidence ($\theta = 0$), according to Eq. (21), we obtain an equation directly for $\delta_y$, with which we have to express the scattered fields,

$$\frac{d^2\delta_y}{d\eta^2} = \frac{e}{m}F(\eta) - \Gamma\frac{(d\delta_y/d\eta)}{\sqrt{1 + (d\delta_y/d\eta)^2/c^2 + (d\delta_y/d\eta)^4/4c^4}}. \qquad (25)$$

We note that both Eq. (24) and Eq. (25), in fact, are first-order differential equations for $d\delta_\perp/d\eta$.

By now, the form of the incoming pulse $F(\eta)$ has not been specified, it can be of arbitrary shape. As is seen from Eqs.(11) and (12), both the reflected and the transmitted signal contain the the unknown term $d\delta_y/dt'$. Due to Eqs. (23), (24) or (25), if once we know $d\delta_y/d\eta$, then the Fourier components of this unknown quantity $d\delta_y/dt'$ can be expressed as

$$\frac{d\delta_y}{dt'}(\omega) = \int_{-\infty}^{+\infty} d\eta\,\frac{d\delta_y}{d\eta}\exp\{i\omega[\eta + \delta_\parallel(\eta)/c]\}, \qquad (26)$$

where we have taken into account the relation $\eta = t' - \vec{n}\cdot\vec{\delta}/c = t' - \delta_\parallel/c$. In this way the solution of the scattering problem is reduced to the solution of the (non-linear) differential equation Eq. (24) (or, in case of perpendicular incidence, Eq. (25)).

### 3. High-harmonic generation on a plasma layer in the moderate relativistic regime

In the present section we consider moderately relativistic motions in the plasma layer, hence we approximate the square root by 1 and neglect the second term in the nominator in Eq.(24). The resulting equation for $d\delta_\perp / d\eta$ yields

$$\frac{d^2\delta_\perp}{d\eta^2} = \frac{e}{m} F(\eta) - (\Gamma \cos\theta) \frac{d\delta_\perp}{d\eta}, \qquad (27)$$

which is formally coincides with Newton equation, Eq. (13), suitable for the non-relativistic description. Henceforth, for simplicity, we denote $\eta$ by $t$. Assuming an impinging Gaussian pulse

$$F(t) = F_0 \exp(-t^2 / 2\tau^2) \cos(\omega_0 t + \varphi) \qquad (28)$$

of amplitude $F_0$, carrier frequency $\omega_0$, pulse width $\tau = \tau_L / 2$ (where $\tau_L$ is the width of the intensity), and carrier-envelope phase difference $\varphi$, Eq. (27) can be approximately solved, yielding

$$d\delta_\perp / dt \approx (eF_0 / m\omega_0)\left(1/\sqrt{1 + \Gamma^2 \cos^2\theta / \omega_0^2}\right) \exp(-t^2 / 2\tau^2) \sin(\omega_0 t + \varphi + \alpha), \qquad (29)$$

where the additional phase $\alpha$ is defined by the relation

$$\sin\alpha = \frac{(\Gamma/\omega_0)\cos\theta}{\sqrt{1 + (\Gamma/\omega_0)^2 \cos^2\theta}}. \qquad (30)$$

The phase term $\delta_\parallel = \vec{n} \cdot \vec{\delta}$ appearing in the exponential in Eq. (26) can be calculated from Eq. (29) by using Eq. (22) to yield

$$\begin{aligned}
\delta_\parallel / c &\approx \frac{1}{2}\left(\frac{eF_0}{mc\omega_0}\right)^2 \frac{1}{1 + (\Gamma/\omega_0)\cos^2\theta} \\
&\times \left\{\frac{1}{2}\int_{-\infty}^{t} dx \exp(-x^2/\tau^2) - \frac{1}{4\omega_0}\exp(-t^2/\tau^2)\sin[2(\omega_0 t + \varphi + \alpha)]\right\}
\end{aligned} \qquad (31)$$

According to Eq. (23), on the basis of Eq. (29), $d\delta_y / dt$ can be expressed as

$$\begin{aligned}
d\delta_y / cdt &\approx 2\beta \cos\theta \exp(-t^2/2\tau^2)\sin(\omega_0 t + \varphi + \alpha) \\
&+ \beta^2 \sin\theta \exp(-t^2/\tau^2)\{1 - \cos[2(\omega_0 t + \varphi + \alpha)]\}
\end{aligned}, \qquad (32)$$

where we have introduced the dimensionless parameter

$$\beta \equiv \frac{1}{2}\left(\frac{eF_0}{mc\omega_0}\right)\frac{1}{\sqrt{1 + (\Gamma/\omega_0)^2 \cos^2\theta}} = \frac{\mu}{2}\frac{1}{\sqrt{1 + (\Gamma/\omega_0)^2 \cos^2\theta}}, \qquad (33)$$

where $\mu \equiv eF_0 / mc\omega_0$ is the dimensionless intensity parameter usually appearing in strong field phenomena. Its numerical value can be calculated according to the formula $\mu = 10^{-9}\sqrt{I}/E_{ph}$, where $I$ is the peak laser intensity measured in $W/cm^2$ and $E_{ph}$ is the average laser photon energy measured in $eV$. Another well-known relation is $\mu^2 = 10^{-18} I\lambda^2$, where the wavelength is measured in microns.

The main problem in calculating the scattered (e. g. the reflected) spectrum is the presence of the time integral in Eq. (31), which is – according to Eq. (26) – is present in the exponential of the Fourier integral. In order to get rid of this difficulty, we approximate this time integral in the following manner

$$\int_{-\infty}^{t} dx \exp(-x^2/\tau^2) \approx \begin{cases} 0, & -\infty \leq t \leq -\tau\sqrt{\pi}/2 \\ \tau\sqrt{\pi}/2 + t, & -\tau\sqrt{\pi}/2 \leq t \leq \tau\sqrt{\pi}/2 \\ \tau\sqrt{\pi}, & \tau\sqrt{\pi}/2 \leq t \leq +\infty \end{cases}. \quad (34)$$

We have numerically checked that the right hand side of Eq.(34) quite reasonably approximates the integral on the left hand side. We also note that for $\tau \to \infty$ (which corresponds to a very long laser pulse), only the second range gives a contribution. Accordingly, we split the time integral in Eq. (26) into three parts

$$\frac{d\delta_y}{dt'}(\omega) = \int dt \frac{d\delta_y(t)}{dt} \exp\{i\omega[t + \delta_\parallel(t)/c]\} \approx$$
$$\int_{-\infty}^{-\tau\sqrt{\pi}/2} B(t)e^{i\omega t} + e^{i\beta^2\tau\sqrt{\pi}\omega/2} \int_{-\tau\sqrt{\pi}/2}^{+\tau\sqrt{\pi}/2} B(t)e^{i(1+\beta^2)\omega t} + e^{i\beta^2\tau\sqrt{\pi}\omega} \int_{+\tau\sqrt{\pi}/2}^{+\infty} B(t)e^{i\omega t} \quad , \quad (35)$$

where

$$B(t) \approx \sum_n J_n(\beta^2\omega/2\omega_0)\{2\beta\cos\theta[e^{-i(2n-1)(\varphi+\alpha)}e^{-(n+1/2)t^2/\tau^2}e^{-i(2n-1)\omega_0 t} -$$
$$e^{-i(2n+1)(\varphi+\alpha)}e^{-(n+1/2)t^2/\tau^2}e^{-i(2n+1)\omega_0 t}]/2i + \beta^2\sin\theta[2e^{-i2n(\varphi+\alpha)}e^{-(n+1)t^2/\tau^2}e^{-i2n\omega_0 t} - \quad . \quad (36)$$
$$e^{-i(2n-2)(\varphi+\alpha)}e^{-(n+1)t^2/\tau^2}e^{-i(2n-2)\omega_0 t} - e^{-i(2n+2)(\varphi+\alpha)}e^{-(n+1)t^2/\tau^2}e^{-i(2n+2)\omega_0 t}]/2\}$$

In obtaining Eq. (36) we have used the Jacobi-Anger formula, the generating formula for the Bessel functions, $e^{-iz\sin\psi} = \sum_n J_n(z)e^{-in\psi}$, moreover we have made the approximation $J_n[(\beta^2\omega/2\omega_0)e^{-t^2/\tau^2}] \approx J_n(\beta^2\omega/2\omega_0)e^{-nt^2/\tau^2}$, which means that – concerning the time dependence – we keep only the leading term in the power expansion of the Bessel functions $J_n$. Within these approximations the integrals in Eq. (35) can be calculated by using the formulae 3.322.1-2 of Gradshteyn and Ryzhik [24].

We have checked that for relatively long pulses, say, for a 10-cycle pulse, at a relativistic intensity $2 \times 10^{19} W/cm^2$, only the second term in Eq. (35) contributes considerably to the spectrum whose maxima correspond to the *intensity-dependend frequency-shifted harmonics* of frequencies $\omega_n = n\omega_0/(1+\beta^2)$, where $\beta$ was defined in Eq. (33). We have found that the spectrum has a very *sharp cut-off determined by the critical index* depending here on the factor $\beta^2/2$ in the argument of the Bessel function. We can borrow a formula for this critical index $n_c$ from the theory of synchrotron radiation (see e.g. Jackson [25]); $n_c = 3/(1-\beta^4/4)^{3/2}$, whose numerical value of is approximately 78 in the case of 45° of incidence. The critical normalized frequency becomes $v_c = \omega_c/\omega_0 = n_c/(1+\beta^2) \approx 28$ which is in agreement whith what we have seen from our numerical calculations. Of course, for this estimate we have to assume that $\beta^2/2 < 1$, namely that the mentioned factor in the argument of the Bessel function is close to, but smaller than one.

To show the spectra for short, 2-cycle pulses we assume the electron density $n_e = 10^{21} cm^{-3}$ and thickness $l_2 = \lambda/100 = 8nm$ for the plasma layer as above, but we take a "moderate" intensity, namely $2 \times 10^{18} W/cm^2$, so one order of magnitude smaller as in the previous example. Then we obtain $v_n = (\omega_n/\omega_0) = n/(1+\beta^2) = \{0.84, 1.68, 2.51, 3.35\}$ for the first four harmonics $n = 1, 2, 3, 4$, where the parameter $\beta$ defined in Eq. (33) is proportional with the usual intensity parameter $\mu = 10^{-9} I^{1/2} \lambda$.

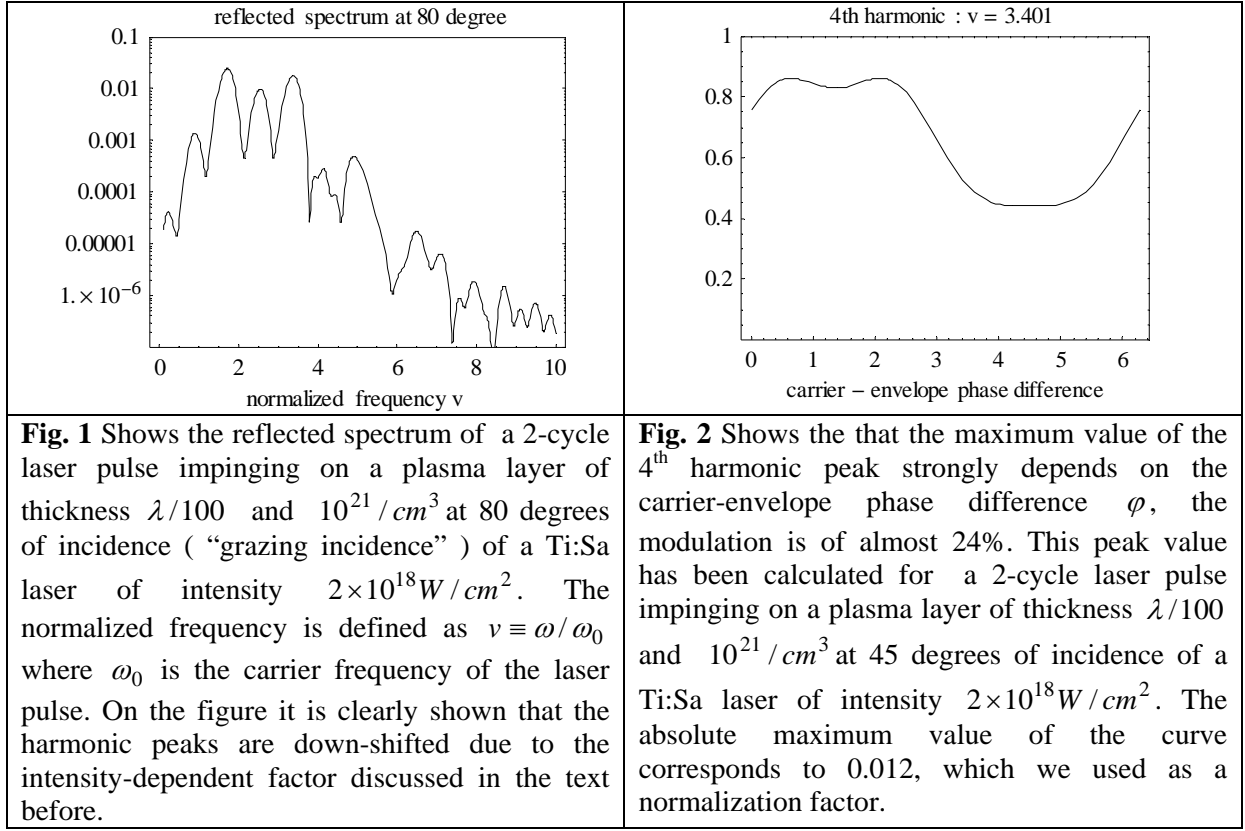

**Fig. 1** Shows the reflected spectrum of a 2-cycle laser pulse impinging on a plasma layer of thickness $\lambda/100$ and $10^{21}/cm^3$ at 80 degrees of incidence ( "grazing incidence" ) of a Ti:Sa laser of intensity $2 \times 10^{18} W/cm^2$. The normalized frequency is defined as $v \equiv \omega/\omega_0$ where $\omega_0$ is the carrier frequency of the laser pulse. On the figure it is clearly shown that the harmonic peaks are down-shifted due to the intensity-dependent factor discussed in the text before.

**Fig. 2** Shows the that the maximum value of the 4$^{th}$ harmonic peak strongly depends on the carrier-envelope phase difference $\varphi$, the modulation is of almost 24%. This peak value has been calculated for a 2-cycle laser pulse impinging on a plasma layer of thickness $\lambda/100$ and $10^{21}/cm^3$ at 45 degrees of incidence of a Ti:Sa laser of intensity $2 \times 10^{18} W/cm^2$. The absolute maximum value of the curve corresponds to 0.012, which we used as a normalization factor.

On **Fig. 1** we see a typical spectrum of the reflected signal at grazing incidence, where the even harmonics dominate. This is understandable from Eqs. (35) and (36) where the angular prefactors $\cos\theta$ and $\sin\theta$ multiply the odd and the even contributions, respectively. In the next example we illustrate on **Fig. 2** that the maximum value of the 4$^{th}$ harmonic peak strongly depends on the carrier-envelope phase difference $\varphi$, the modulation (which is defined as $M = (I_{max} - I_{min})/(I_{max} + I_{min})$) is of almost 25%. We mention that we have checked this modulation $M$ for the other harmonic peaks, but, interestingly, we found it very small ($M < 1 - 2\%$). Similarily for the integrated spectrum the value of $M$ is also very small, a couple of per cents.

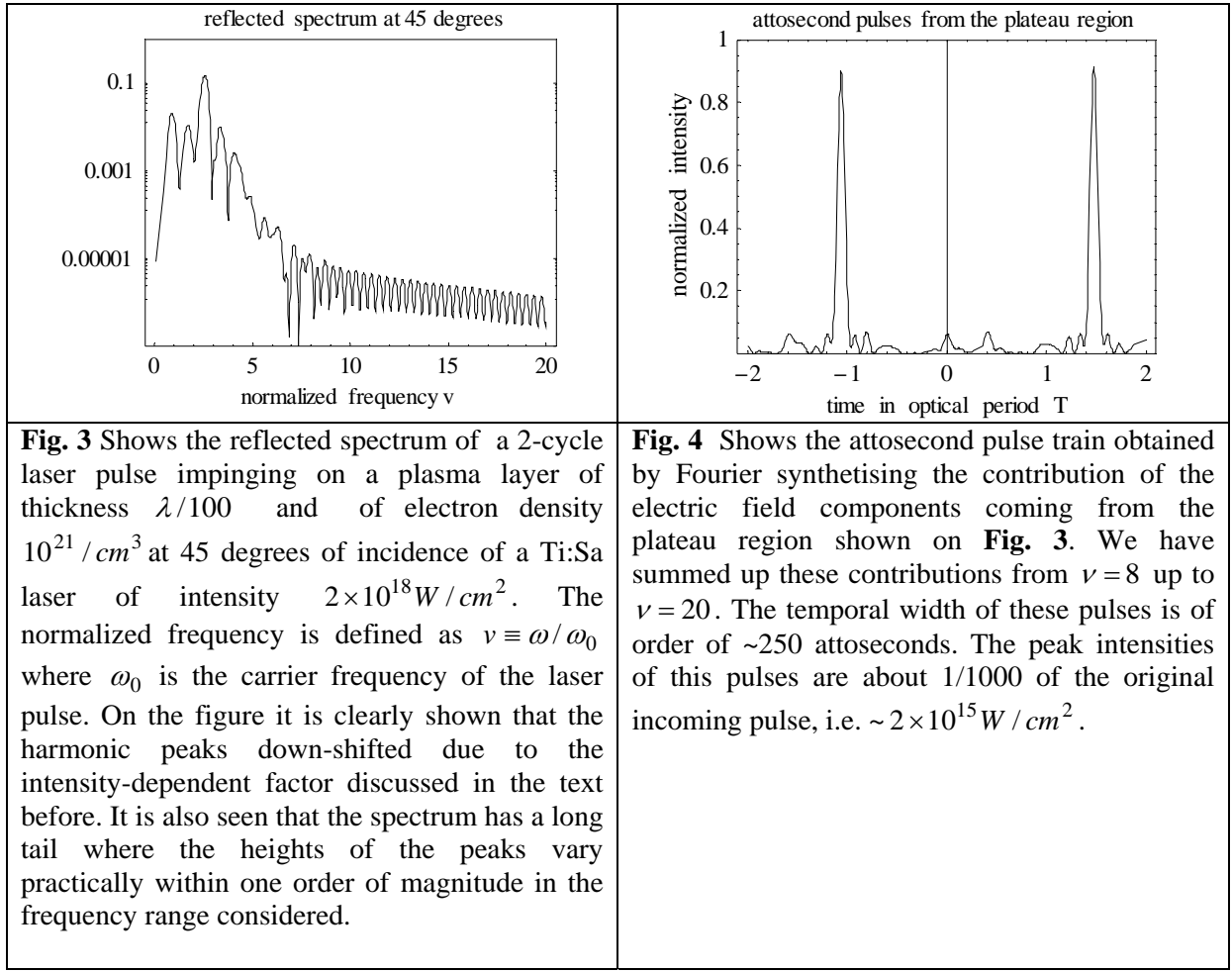

**Fig. 3** Shows the reflected spectrum of a 2-cycle laser pulse impinging on a plasma layer of thickness $\lambda/100$ and of electron density $10^{21}/cm^3$ at 45 degrees of incidence of a Ti:Sa laser of intensity $2\times 10^{18} W/cm^2$. The normalized frequency is defined as $\nu \equiv \omega/\omega_0$ where $\omega_0$ is the carrier frequency of the laser pulse. On the figure it is clearly shown that the harmonic peaks down-shifted due to the intensity-dependent factor discussed in the text before. It is also seen that the spectrum has a long tail where the heights of the peaks vary practically within one order of magnitude in the frequency range considered.

**Fig. 4** Shows the attosecond pulse train obtained by Fourier synthetising the contribution of the electric field components coming from the plateau region shown on **Fig. 3**. We have summed up these contributions from $\nu = 8$ up to $\nu = 20$. The temporal width of these pulses is of order of ~250 attoseconds. The peak intensities of this pulses are about 1/1000 of the original incoming pulse, i.e. $\sim 2\times 10^{15} W/cm^2$.

The presence of the "plateau" shown on **Fig. 3** suggests that by Fourier synthesis of this part of the spectrum results in the appearance of attosecond pulses. Note that the spacing of the individual peaks in the plateau region is much smaller than the central frequency, thus this plateau is almost a quasi-continuum. The result of the Fourier syntheses is shown on **Fig. 4**, where we see a part of an attosecond pulse train in the reflected signal.

By closing the present letter, we can conclude from our analysis, that, on one hand, certain parts of the scattered spectrum of a few-cycle laser pulse impinging on a plasma layer is CE phase-sensitive, and, on the other hand, the Fourier components in the plateau region are in phase, and coherently add up to yield high-intensity attosecond UV pulse trains. Moreover, we have have given an analytic expression for the critical index characterising the cut-off region of the harmonic spectra, and we have proven that in general there is an intensity-dependent frequency down shift of the harmonic components generated on the plasma layer, analogous to that had been found earlier for nonlinear Thomson scattering on a single electron.

**Acknowledgements.** This work has been supported by the Hungarian National Scientific Research Foundation, OTKA, Grant No. T048324.

Sándor Varró: Scattering of an few-cycle laser pulse by a plasma layer… 11[3] M. Hentschel, R. Kienberger, Ch. Spielmann, G. A. Reider, N. Milosevic, Th. Brabec, P. Corkum, U. Heinzmann, M. Drescher and F. Krausz, Nature **414**, 509-513 (2001)

[4] M. Drescher, M. Hentschel, R. Kienberger, M. Üiberacker, V. S. Yakovlev, A. Scrinzi, Th. Westerwalbesloh, U. Kleineberg, U. Heinzmann and F. Krausz, Nature **419**, 803-807 (2002)

[5] S. Chelkowski, A. D. Bandrauk and A. Apolonskiy, Phys. Rev. A **65**, 061802(R)(1-4) (2002)

[6] D. B. Milosevic, G. G. Paulus and W. Becker, Phys. Rev. Lett. **89**, 15300(1-4) (2002)

[7] D. B. Milosevic, G. G. Paulus and W. Becker, Optics Express **11**, 1418-1429 (2003)

[8] A. Baltuska, Th. Udem, M. Üiberacker, M. Hentschel, E. Goulleimakis, Ch. Gohle, R. Holtzwarth, V. S. Yakovlev, A. Scrinzi, T. W. Hänsch, F. Krausz, Nature **421**, 611-615 (2003)

[9] S. Witte, R. T. Zinkstok, W. Hogerworst and K. S. E. Eikema, Appl. Phys. B **78**, 5-12 (2004)

[10] Ch. Lemell, X.-M. Tong, F. Krausz and J. Burgdörfer, Phys. Rev. Lett. **90**, 076403(1-4) (2003)

[11] A. Apolonskiy, P. Dombi, G. G. Paulus, M. Kakehata, R. Holtzwarth, Th. Udem, Ch. Lemell, K. Torizuka, J. Burgdörfer, T. W. Hänsch, F. Krausz, Phys. Rev. Lett. **92**, 073902(1-4) (2004)

[12] P. Dombi, A. Apolonskiy, Ch. Lemell, G. G. Paulus, M. Kakehata, R. Holtzwarth, Th. Udem, K. Torizuka, J. Burgdörfer, T. W. Hänsch and F. Krausz, New J. Phys. **6**, 39-48 (2004)

[13] T. M. Fortier, P. A. Roos, D. J. Jones and S. T. Cundiff, Phys. Rev. Lett. **92**, 147403(1-4) (2004)

[14] T. Nakajima and Sh. Watanabe, Phys. Rev. Lett. **96**, 213001(1-4) (2006)

[15] H. Fearn and W. E. Lamb, Jr., Phys. Rev. A **43**, 2124-2128 (1991)

[16] T. Ristow, Diplom work (unpublished). (Rheinisch-Westfälische Technische Hochschule, Aachen, Germany, 2004)

[17] S. Varró, Laser Phys. Lett. **1**, 42-45 (2004)

[18] A. Sommerfeld, Ann. der Physik, **46**, 721-747 (1915)

[19] R. Lichters, J. Meyer-ter-Vehn and A. Pukhov, Phys. Plasmas, **3**, 3425-3432 (1996)

[20] S. Kiselev, A. Pukhov and I. Kostyukov, Phys. Rev. Lett. **93**, 135004(1-4) (2004)

[21] F. Quéré, C. Thaury, P. Monot, S. Dobosz, Ph. Martin, J.-P. Geindre and P. Audebert, Phys. Rev. Lett. **96**, 125004(1-4) (2006)

[22] A. Pukhov and J. Meyer-ter-Vehn, Appl. Phys. B 74, 355-361 (2002)

[23] A. Hidding, K.-A. Amthor, B. Liesfeld, H. Schwoerer, S. Karsch, M. Geissler, L. Veisz, K. Schmid, J. G. Gallacher, S. P. Jamison, D. Jaroszynski, G. Pretzler and R. Sauerbrey, Phys. Rev. Lett. **96**, 105004(1-4) (2006)

[24] I. S. Gradshteyn and I. M. Ryzhik, Table of Integrals, Series and Products (Academic Press, New York,1980)

[25] J. D. Jackson, Classical Electrodynamics (John Wiley & Sons, Inc., New York, 1962)